# Scalable Quantum Reversible BCD Adder Architectures with Enhanced Speed and Reduced Quantum Cost for Next-Generation Computing

Negin Mashayekhi[1], Mohammad Reza Reshadinezhad[1,*], Antonio Rubio[2,*], Shekoofeh Moghimi[1]

[1] Faculty of Computer Engineering Department, University of Isfahan, Isfahan, Iran.

[2] Electrical Engineering Department, Polytechnic University of Catalonia, Barcelona, Spain.

E-mail: n.mashayekhi@eng.ui.ac.ir, m.reshadinezhad@eng.ui.ac.ir, antonio.rubio@upc.edu, shokufemoghimi@yahoo.com

**Keywords**: Quantum Computing, Reversible BCD Adder, Carry Skip Technique, Quantum Delay, Quantum Cost.

*Abstract*: The quantum and reversible paradigm merges the principles of quantum mechanics and reversible computation to enable information-preserving processing. It supports next-generation computing architectures that provide improved scalability and enhanced computational efficiency. Within these architectures, the decimal adder is a key arithmetic component, particularly for Binary Coded Decimal (BCD) operations widely used in financial and commercial systems. However, most reversible BCD adders focus primarily on quantum and reversible metrics, overlooking the critical influence of delay, which makes balanced optimization a significant challenge. This paper presents two reversible BCD adder designs optimized for both delay and quantum cost. One design integrates the decimal carry-skip technique to improve the overall delay. Using reversible logic gates, the proposed architectures efficiently perform BCD addition and implement the required correction logic while maintaining full reversibility. Evaluation results indicate that the proposed designs surpass existing reversible BCD adders, achieving best-case average improvements of 85.12% in delay and 30.75% in quantum cost. These advancements demonstrate the potential of the proposed adders for integration into future quantum-based arithmetic units and scalable reversible computing systems. Moreover, analysis of real banking transaction data underscores the practical importance of BCD addition and its widespread use in accurate and efficient monetary computations.

## 1. Introduction

Reversible computing represents a transformative paradigm that preserves information integrity throughout computation, thereby addressing the intrinsic thermodynamic limitations of conventional computing systems. By enforcing a one-to-one correspondence between input and output vectors, it not only mitigates the energy dissipation limits articulated by Landauer's [1] principle but also establishes a fundamental basis for quantum circuit design, where all physical operations are inherently reversible. Landauer noted that classical gate-based logic computation results in information loss, leading to energy or heat loss of at least $KTln2$ Joules per bit, where $K$ is Boltzmann's constant and $T$ is the absolute temperature. In large circuits, the energy loss scales with the number of gate counts or bits. Bennett [2] demonstrated that the energy dissipation in conventional computation can be avoided by employing reversible computational mechanisms, which in turn inspired the creation of reversible logic gates. With the continued advancement of quantum computing, the design of reversible logic circuits has become increasingly important, forming the essential foundation for quantum information processing. This intrinsic connection underscores the critical role of reversible logic development as a necessary step toward practical quantum computation [3–5].

Among the fundamental arithmetic operations, addition is important. For financial applications, where decimal arithmetic is essential to avoid rounding errors in monetary calculations, Binary-Coded Decimal (BCD) adders are a crucial component [6- 9]. However, the conventional BCD adder is designed with delay and cost overhead. A primary challenge is their delay, stemming from the use of a preliminary 4-bit binary adder followed by a correction circuit, which must determine if the result is a valid BCD digit.

Driven by the need for high-performance computing, numerous studies have emphasized the design of a reversible circuit. Various designs and methods have been proposed to improve reversible adders in terms of Quantum Cost (QC), Garbage Output (GO), and Constant Input (CI). Existing works [10-15] in this domain frequently employed reversible gates such as the Feynman Gate (FG) [16], Toffoli Gate (TG) [17], Peres Gate (PG) [18], Fredkin Gate (FRG) [19], HNG gate [20], ZPLG gate [21], MTSG gate [22], and BJN gate [23]. While prior works have designed the reversible BCD adders, their focus has been predominantly on minimizing quantum cost. Consequently, these designs exhibit propagation delay, which limits their practicality for large-scale, high-speed applications.

This work focuses on the optimization of the two primary quantum metrics: quantum cost and delay. We aim to achieve a superior trade-off between these parameters. The article presents two new proposed designs. The first utilizes existing reversible gates to construct an efficient Six Correction Logic (SCL) module and proposes a new decimal full adder. The second design utilizes the first proposed component and presents a new structure for the BCD addition circuit, incorporating a speedup technique. The main novelty of this work is in the proposed innovative BCD architecture that combines the proposed reversible-based BCD adder with a carry skip technique. By integrating these two efficient structures, our design achieves a significant reduction in both critical path delay and overall quantum cost, ultimately delivering a highly efficient and balanced reversible BCD adder suitable for the demanding requirements of future quantum and financial computing systems.

The rest of this research article is organized as follows: Section 2 covers the background of reversible logic, quantum costs, and the BCD adder. Section 3 provides a brief review of previous works. Section 4 presents the proposed designs. Section 5 is dedicated to evaluation and comparison with prior designs. Section 6 concludes the article.

## 2. Background

2.1. Reversible Concept

The primitive logical unit of a reversible circuit is known as a reversible gate. For a gate to be reversible, it must possess specific characteristics that set it apart from other gates. These characteristics include [24]:

1. The number of inputs and outputs of the circuit should be equal.
2. There must be a one-to-one correspondence between inputs and outputs, ensuring that each input maps to a unique output. This allows for the derivation of a unique output pattern from each input pattern and enables the input pattern to be determined from its corresponding output.

A reversible circuit is constructed solely using reversible gates and follows distinct and challenging rules that differ significantly from those of conventional circuits in classical logic. Notably, reversible circuit design prohibits the use of fan-out, meaning each input and output signal is used only once. Additionally, reversible logic circuits do not permit feedback or return loops to previous stages, preventing any output line from being reconnected to former stages. Reversible circuits have certain figure of merit, such as QC, CI, and GO, which are explained in Section 5.

*Quantum Cost (QC)*

Quantum cost is a crucial factor for evaluating logic gates in reversible circuits, defined as the total of the quantum costs for all reversible gates used in the circuit. Gates sized at 1×1 do not contribute to the quantum cost. Each 2×2 gate adds a quantum cost of 1, with larger gates calculated accordingly. The square root of a NOT gate is represented as V, and its Hermitian counterpart as V+. The quantum cost of a reversible gate is calculated by summing the number of elementary quantum operations, namely the V, $V^+$, and CNOT gates [25, 26].

*Constant Inputs*

Constant inputs in a reversible gate are set to either 0 or 1 and are crucial for maintaining reversibility in a function. Reducing the number of constant inputs is key to enhancing reversible circuit design [27]. In reversible circuits, besides the primary inputs, certain constant input bits, known as ancilla inputs, are used to implement various logic functions. These ancilla inputs are considered overhead and need to be minimized [11].

*Garbage Outputs*

In reversible circuits, some outputs are unused but are essential for preserving circuit reversibility. Reducing the number of garbage outputs is crucial for designing efficient reversible circuits [4]. Garbage outputs, which are necessary to maintain reversibility but do not contribute to useful computations, are considered overhead and need to be minimized in reversible designs [11].

2.2. Reversible gate

Several reversible logic gates have been introduced in the literature. Table 1 presents the Quantum Cost (QC) and the delay for the reversible gates used in our designs, as well as in some previous designs.

Table 1. Reversible gates such as FG, PG, M-F, HNG, and BJN, and their quantum circuits and metrics

| Reversible gates | Parameters | Quantum- Circuit |
|---|---|---|
| 1. FG [16]<br>P = A<br>Q = A ⊕ B | QC = 1<br>Delay = 1Δ | A •——— P<br>B ⊕——— Q |
| 3. PG [18]<br>P = A<br>Q = A ⊕ B<br>R = AB ⊕ C | QC = 4<br>Delay = 4Δ | |
| 4. M-F [28]<br>P = A<br>Q = $\bar{A}$B ⊕ A$\bar{C}$<br>R = AB ⊕ $\bar{A}$C | QC = 4<br>Delay = 3Δ | |
| 5. HNG [20]<br>P = A<br>Q = B<br>R = A ⊕ B ⊕ C<br>S = (A ⊕ B) C ⊕ AB ⊕ D | QC = 6<br>Delay = 5Δ | |
| 6. BJN [23]<br>P = A<br>Q = B<br>R = (A + B) ⊕ C | QC = 5<br>Delay = 4Δ | |

### 2.3. BCD adder

When two BCD digits are added, considering carry from the preceding lower-order digit (worst case: 9 + 9 + 1), the maximum possible sum is 19. The BCD adder circuit consists of three main functional stages, described as follows:

1. Initial Addition: Two 4-bit BCD operands are added using a conventional 4-bit binary adder. Depending on the input values, the resulting sum ($S_i$) may range from 0 (0000) to 18 (10010) in binary form.
2. Invalid Result Detection: After binary addition, the output is examined for validity, and a decimal carry is generated through the *Six Correction Logic (SCL)* block. In a valid BCD representation, binary results must lie between "0000" (0) and "1001" (9). An invalid result occurs when either:
   o The 4-bit sum exceeds "1001" (decimal 9), or
   o A carry-out ($C_{out} = 1$) is generated from the adder.
3. Correction Mechanism: For invalid results, a correction of 0110 (decimal 6) is added to the binary output. This adjustment converts the invalid binary sum into a valid BCD output while ensuring correct carry propagation for multi-digit additions.

In **Figure 1**, the block diagram of the Decimal Full Adder (DFA) is illustrated, where $dA_j$, $dB_j$, and $dS_j$ Denote BCD digits, and $dC_{i+1}$ Represents the decimal *carry-out* bit. The DFA block at the position j (where j indicates the position of each of two digits) computes $dA_j + dB_j + C_j = 10dC_{(j+1)} + dS_j$ ($C_0 = 0$). The DFA block employs a 4-bit binary adder to calculate the resulting sum digit, and corrects it to $S_i - 10$ (or $|S_i + 6|$) if $S_i > 10$ and $dC_{j+1} = C_4 + S_3(S_2 + S_1)$.

(Where $dA_j = (a_3a_2a_1a_0)_j$, $dB_j = (b_3b_2b_1b_0)_j$, and $dS_j = (S_3S_2S_1S_0)_j$ are BCD digits that contain 4-bit and $C_4$ is the binary carry output of a 4-bit binary adder. Equation 1 indicates the function of the SCL block that computes the decimal carry output.

$$dC_{j+1} = C4 \oplus S_3(S_2 + S_1) \qquad (1)$$

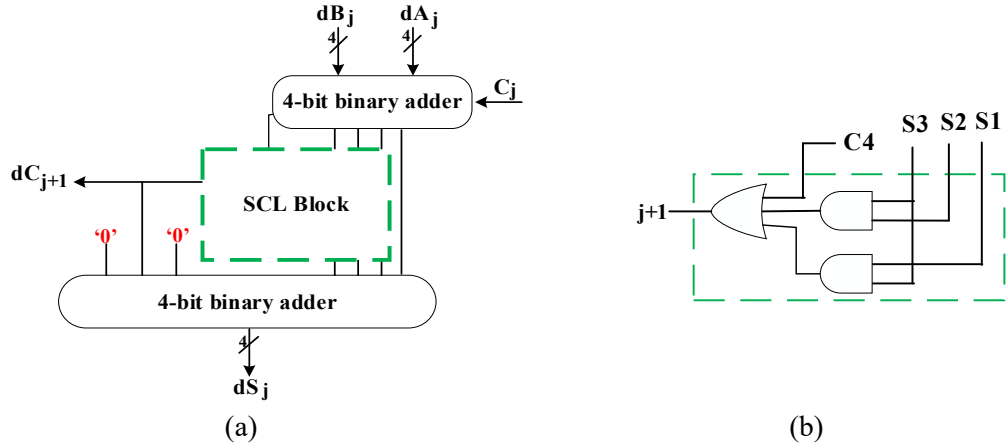

Figure 1. (a) Block diagram of DFA. (b) Details of an SCL Block [8].

2.4. Carry Skip Adder

The Carry Skip (CSK) adder enhances computational speed by allowing the carry signal to bypass specific bit positions where propagation is not required. It achieves this by detecting conditions in which the corresponding bits of the input operands differ, thereby eliminating unnecessary carry propagation through intermediate stages. In this approach, it is verified whether a carry is passed on to the subsequent block or not. This adder design provides low delay, making it ideal for high-speed arithmetic tasks. **Figure 2** depicts a case of CSK implementation, where FA stands for a binary Full Adder and P indicates the propagating signal. In binary mode, the carry skip technique is applied, where $p_i = a_i \oplus b_i = 1$ (where $a_i$ and $b_i$ are input bits and $p_i$ called the propagate signal). An *n*-bit adder is partitioned into several blocks, where each block consists of a set of consecutive stages implemented using a basic RCA structure. Within each block, a group carry skip signal (P) is generated. This signal is asserted (i.e., set to one) when the propagate condition is met. If $p_i = 1$ is satisfied for every bit in the group. When this condition is satisfied, the carry-in signal can bypass the entire block and be directly transferred to the carry-out, without being influenced by intermediate carry generation or propagation calculation within the group. In this architecture, a skip signal P is generated for each group of full adders. For a block spanning from the stage $j$ to $i$, the skip signal is defined as the logical AND of the individual propagate signals, $P = p_jp_{j+1} \dots p_i$ [6, 8].

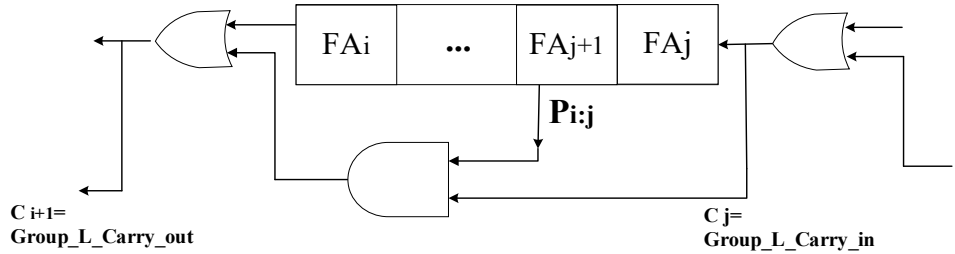

**Figure 2**. Block diagram of the L-th group of a carry-skip adder, including bits $j, j + 1, \ldots, i$..

## 3. Related Works

Extensive research has been conducted on reversible computation, particularly in the areas of logic synthesis and optimization. There is significant state-of-the-art work on reversible BCD adders and sequential circuits. Numerous studies, such as those in [10-15, 29-33], have explored the synthesis of reversible logic circuits. Researchers have focused on optimizing these circuits by reducing quantum cost and the number of constant inputs and garbage outputs.

In 2020, Thabah proposed two fast designs for reversible BCD adders. The first design utilized reversible gates like BJN and TG in the correction logic of the BCD adder, while the second design introduced a new BCD adder using only basic reversible gates such as PG and TG. According to the results, the first design achieved better outcomes in terms of figure of merits [10].

Thapliyal et al. [11] proposed two innovative reversible ripple carry adder architectures: (i) a design that functions without an input carry and does not utilize any ancilla input bits, and (ii) one with input carry and lacking ancilla input bits. The proposed reversible ripple carry adder designs feature no ancilla input bits and a limited number of garbage outputs, all while achieving enhanced values for quantum cost. In these designs, both quantum cost and delay are decreased by creating designs grounded in the reversible PG and the TR gate. Based on this adder, innovative designs for the reversible BCD adder are introduced following two approaches: (i) the addition occurs in binary mode, and correction is made to switch to BCD, when necessary, through detection and correction, and (ii) the addition takes place in binary format, with the outcome consistently converted via a binary to BCD converter. These designs are refined for the count of ancilla inputs and the quantity of garbage outputs. Based on the findings, the second proposed design yielded superior results concerning the figure of merit. Haghparast proposed two reversible logic gates, HNFG and HNG, in [12]. The HNFG gate can function as two FGs, making it suitable for creating a single copy of two bits without producing garbage outputs. It can serve as a "Copying Circuit" to enhance fan-out, which is not permitted in reversible circuits. The HNG gate is capable of implementing all Boolean functions and can be used to design optimized adder architectures. This proposed full adder is then utilized in designing a reversible 4-bit parallel adder and a reversible BCD adder circuit. In the other article [15], Haghparast et al. presented new reversible gates with a parity-preserving feature. They proposed a new BCD adder constructed using the newly proposed gates for the binary adder and BCD converter in the correction step. In [13], Islam implemented a new BCD adder that features a quantum cost-efficient design for nanotechnology systems, utilizing PFAG, PG, and HNFG gates. In [14], the study investigated the application of negative control lines and introduced a new reversible gate, NAFA, which was used to develop a novel BCD adder.

## 4. The Proposed Reversible BCD Adders

### 4.1. The Proposed RCA Reversible BCD Adder (*Dec-RCA*)

This work introduces a new design targeting the critical decimal carry-correction stages of the BCD adder. Specifically, it emphasizes the detection of valid decimal carry outputs and the identification and correction of invalid BCD digits through the addition of 'six'. The central contribution is the development of a Proposed Decimal Full Adder (PDFA) block, implemented with efficient quantum-reversible gates to provide an optimized, cost-effective solution for both carry computation and SCL. According to Equation 1, the decimal carry output ($dC_{j+1}$) in the SCL block is computed using the binary carry-out of the binary RCA adder and the sum bits. In this work, efficient reversible gates such as the BJN and PG gates are employed to design a novel SCL block capable of accurately generating the decimal carry output. As shown in Table 1, the BJN gate functions as an OR operator and is utilized to compute $S1 + S2$ with the third input set to '0'. Meanwhile, the PG gate operates as an AND gate, used to generate the decimal carry output, where $(S1 + S2)$, $S3$, and $C4$ serve as inputs, and the result is produced at the third output signal. **Figure 3** illustrates the quantum representations of the proposed SCL block, where S1, S2, and S3 correspond to the partial sum outputs of the binary adder, C4 denotes the carry produced by the 4-bit binary addition, and Cout represents the resulting decimal carry output.

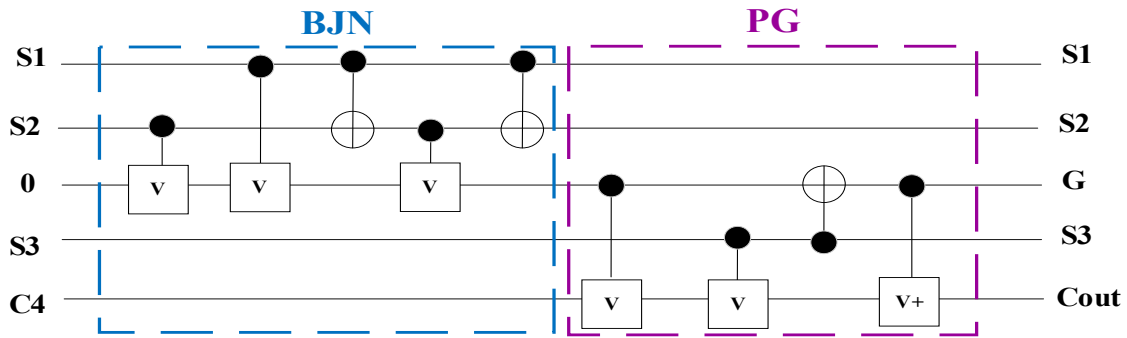

**Figure 3**. Quantum representation of the proposed SCL block by reversible gates.

**Figure 4** demonstrates the proposed PDFA block, which is constructed by the proposed SCL block and generates the value output. The PG gate ultimately produces the final decimal carry value. At the end of the calculation, we use an FG gate to produce a carry copy to propagate to the next digit, and the other gates are for the corrected summation. This proposed block features 2 numbers of CI, 1 number of GO, and has a QC of 10. The BJN gate (within the SCL block) operates in parallel with the final full adder module in the RCA structure without consuming extra time, resulting in a delay of 5Δ for the proposed SCL design.

In the final stage of the circuit, it is necessary to convert the generated output into a valid value. If the result requires correction, the value "0110" must be added to the result at this step. This process ensures that valid outputs remain unchanged and do not require modification. This paper proposes a new structure for the correction section to calculate the addition of binary summation with the value "0110". The second and third outputs are then transferred to a complete adder block (HNG gate), and the Most Significant Bit (MSB) is calculated by the FG gate. The overall design of the Proposed DFA (PDFA) is illustrated in Figure 4. This efficient 1-digit design features 8 numbers of CI, 3 numbers of GO, and a QC of 45.

According to Table 1, the total delay of the entire proposed block is calculated by Equation 2:

$$delay_{PDFA} = delay_{RCA} + delay_{SCl} \quad (2)$$
$$+ delay_{six-correction}$$
$$= 4delay_{HNG}$$
$$+ (delay_{PG} + delay_{FG})$$
$$+ (delay_{PG} + delay_{HNG} + delay_{FG})$$
$$= 20 + 5 + 10 = 35 \, \Delta$$

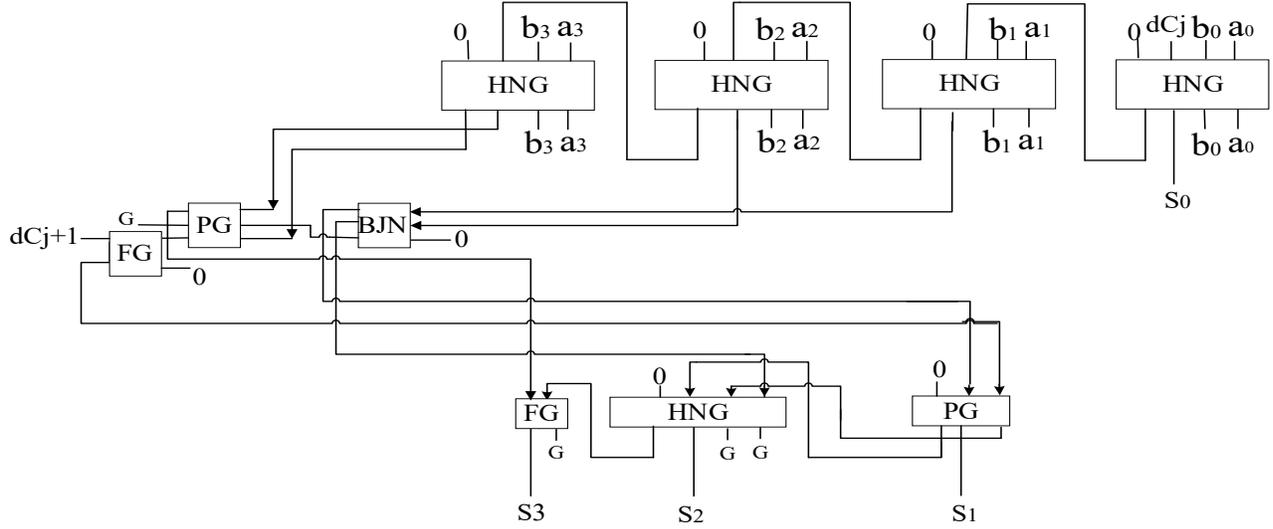

**Figure 4**. Block diagram of the proposed PDFA block.

The proposed RCA-based design, referred to as "*Dec-RCA*" performs decimal addition by cascading the introduced PDFA blocks within a conventional Ripple Carry Adder (RCA) framework, as illustrated in **Figure 5**. Figures 3–5 present the structural details of this first proposed decimal adder using the RCA architecture. The design begins with a 4-bit RCA composed of four binary full-adder stages, where $a_i$, $b_i$, and $c_i$ serve as binary inputs, and $C_4$ and $S_i$ are the resulting binary outputs. Based on Table 1, the HNG gate implements the same logic as a full adder; when its fourth input is set to 0, it produces the sum and carry-out outputs without generating any garbage. Under these conditions, the 4-bit RCA requires 4 constant inputs, achieves a quantum cost of 24, and produces no garbage outputs. Its delay, determined by the HNG gate characteristics, is 20. As depicted in Figure 5, the overall delay of the proposed architecture corresponds to the delay introduced by the cascaded PDFA blocks along the critical path (shown by the dashed line).

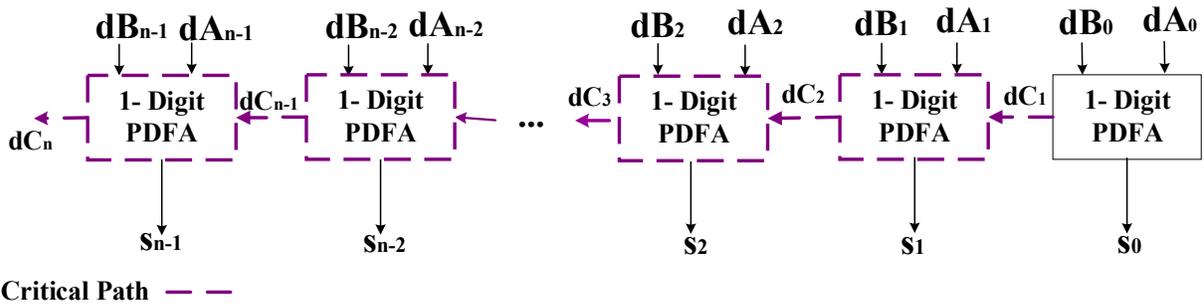

Critical Path — —

**Figure 5**. Block diagram of the proposed RCA reversible BCD adder (*Dec-RCA*) for an n-digit input that is constructed by PDFA, with the critical path indicated by dashed lines.

## 4.2. The Proposed CSK Reversible BCD Adder (*Dec-CSK*)

In the second proposed design, the aim is to modify the *Dec-RCA* design in terms of delay using a speedup technique. The delay of an adder is usually dictated by how effectively the carry chain is handled and by the delay in the critical path of the design. A key emphasis of this part is applying the Carry Skip (CSK) technique (one of the enhanced speed-up techniques) for decimal carry propagation signals within the decimal adder named "*Dec-CSK*". In a decimal adder, the condition for bypassing the carry signal occurs when $dA_j + dB_j = 9$ (where $dA_j$ and $dB_j$ are decimal values) [8]. In the binary calculation, the *propagate* ($p$) and *generate* ($g$) signals are defined as Equations 3 and 4. By considering the condition "$dA_j + dB_j = 9$" and the definition of $p$ and $g$ signals, a function named *decimal propagation* ($P$) that is computed for each digit is defined in Equation 5 [8]. According to Equations 3 to 5, $a_i$ and $b_i$ represent binary variables, and each decimal digit is expressed as a 4-bit binary number denoted by $a_3a_2a_1a_0$. Furthermore, Equation 6 defines the function responsible for generating the decimal carry output, denoted as $G_i$. The proposed *Dec-CSK* design utilizes the decimal propagation and generation signals to compute the correct decimal carry output ($dCj+1$) and proposes it as Equation 7.

$$p_i = a_i \oplus b_i \qquad 0 \leq i \leq 3 \qquad (3)$$

$$g_i = a_i . b_i \qquad 0 \leq i \leq 3 \qquad (4)$$

$$P_j = p_o(p_3 \oplus g_1p_2 \oplus (g_2(\overline{g_1 \oplus p_1}))) \qquad (5)$$

$$G_j = C4 \oplus S_3(S_2 + S_1)) \qquad (6)$$

$$dC_{j+1} = P_j . dC_j + \overline{P_j}(G_j) \qquad (7)$$

**Figure 6** illustrates the *Dec-CSK*, which utilizes the carry skip technique applied to the *Dec-RCA* design to modify the decimal carry propagation delay. We utilize the $P_j$ signals to choose between the decimal carry input and the generated decimal carry output ($G_j$) to ensure accurate decimal carry output. If $P_j$ is equal to 1, it indicates that the carry is propagated from the input to the output. Conversely, if $P_j = 0$, it is determined that carry propagation has not occurred, and the decimal carry out relies on the carry out from the RCA adder. The proposed *Dec-CSK* design utilizes the Modified Fredkin (M-F) [28] reversible gate as a multiplexer and the Double Feynman Gate (DFG) [34] reversible gate to copy the decimal carry output, propagating the carry to the next digit and the next skip block, and to create the correct sum digit. According to Figures 6 and 7, the proposed CSK architecture is designed using the proposed PDFA block as the fundamental arithmetic unit, in conjunction with the previously reversible logic gates. As illustrated in Figure 6, the notations "$dA_j$", "$dB_j$", and "$dS_j$" correspond to the BCD digits of the input operands and their sum, while $dC_{j+1}$ denotes the decimal carry-out produced for each digit. The parameters $P_j$ and $G_j$ represent the decimal propagate and decimal generate signals, respectively. In this configuration, $P_j$ and $G_j$ are generated for each digit using Equations 5 and 6, and are subsequently employed within the CSK framework to accurately compute the decimal carry-out and propagate it to the next digit position.

**Figure 7** provides a more detailed depiction of how the signal. $P_j$ is implemented based on Equation 5. In this design, $P_j$ is produced using PG and M-F reversible gates, which collectively ensure logical consistency and reversibility. The signal $P_j$ functions as a selection control within the skip block, which is composed of an M-F (acting as a reversible multiplexer) and a DFG. The skip block dynamically determines whether to propagate the generated decimal carry signal ($G_j$) or the incoming decimal carry signal ($dC_j$), thereby computing the appropriate carry output ($dC_{j+1}$) for the next digit stage.

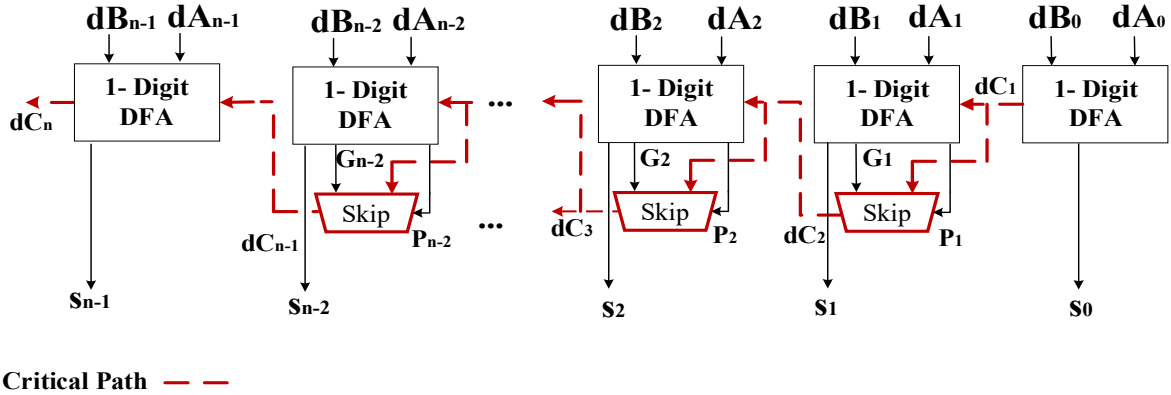

**Figure 6**. Block diagram of the proposed CSK reversible BCD adder (*Dec-CSK*) for an n-digit input, with the critical path indicated by dashed lines.

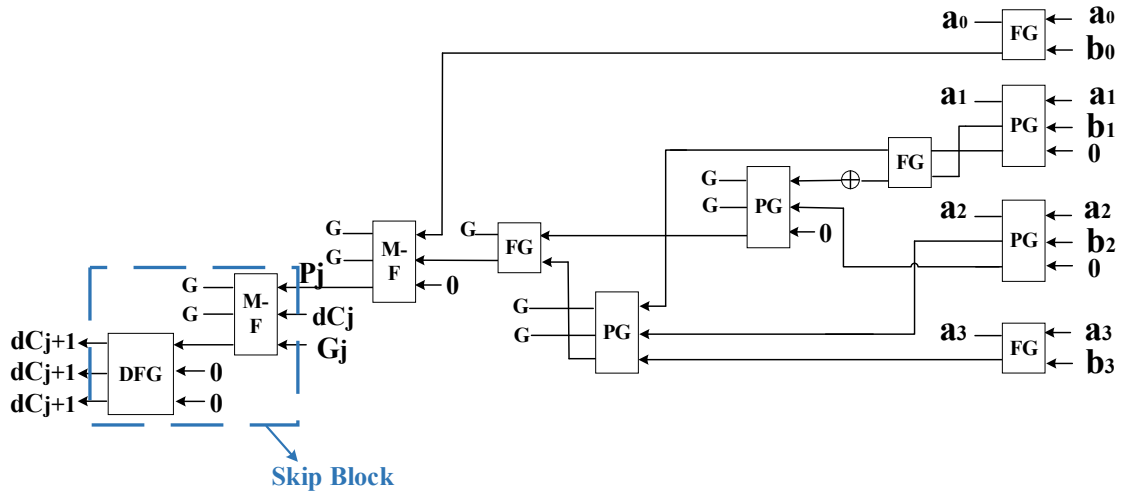

**Figure 7**. The block diagram of the proposed decimal skip signal generator ($P_j$) and skip block in the proposed *Dec-CSK* adder, and indicate the block diagram of the skip by dashed lines.

## 5. Evaluation And Comparison Results

This section presents the simulation and verification of the proposed design using both quantum and classical environments. The quantum behavior of the circuit is first demonstrated through Quirk, an online quantum simulation tool, to illustrate the functional correctness of the proposed PDFA block at the quantum gate level. To further validate the practicality of the design, the overall *Dec-CSK* structure is implemented and tested in ModelSim using a Hardware Description

Language (HDL). This dual-level verification confirms the accuracy and feasibility of the proposed approach in both quantum simulation and hardware implementation contexts. For instance, to validate the correctness and functionality of the proposed design, **Figure 8** presents the simulation results of this block implemented using the Quirk quantum simulator. According to the Quirk online tool's limitation of displaying a maximum of 16 lines, the FG block (which generates a copy of the carry-out) is removed. This modification does not affect the results, and the figure demonstrates the verification of the proposed design. The results presented in Tables 2-5 are obtained from the proposed structure illustrated in Figure 4. All possible input patterns are applied to the input qubits of the circuit. For example, in the Figure. 8, the input signals $A = a_3 a_2 a_1 a_0 = "1001"$ and $B = b_3 b_2 ab_1 ab_0 = "1001"$ are applied to the proposed PDFA circuit, and the corresponding output obtained from the simulation is "11000", which matches the expected circuit behavior. In this simulation, the signals $S_3 S_2 S_1 S_0$ correspond to the summation outputs. The signal $dCj$ represents the decimal carry-in, which in this case holds the value '0'. On the output side of the simulation, the decimal carry-out ($dC_{j+1}$) is generated.

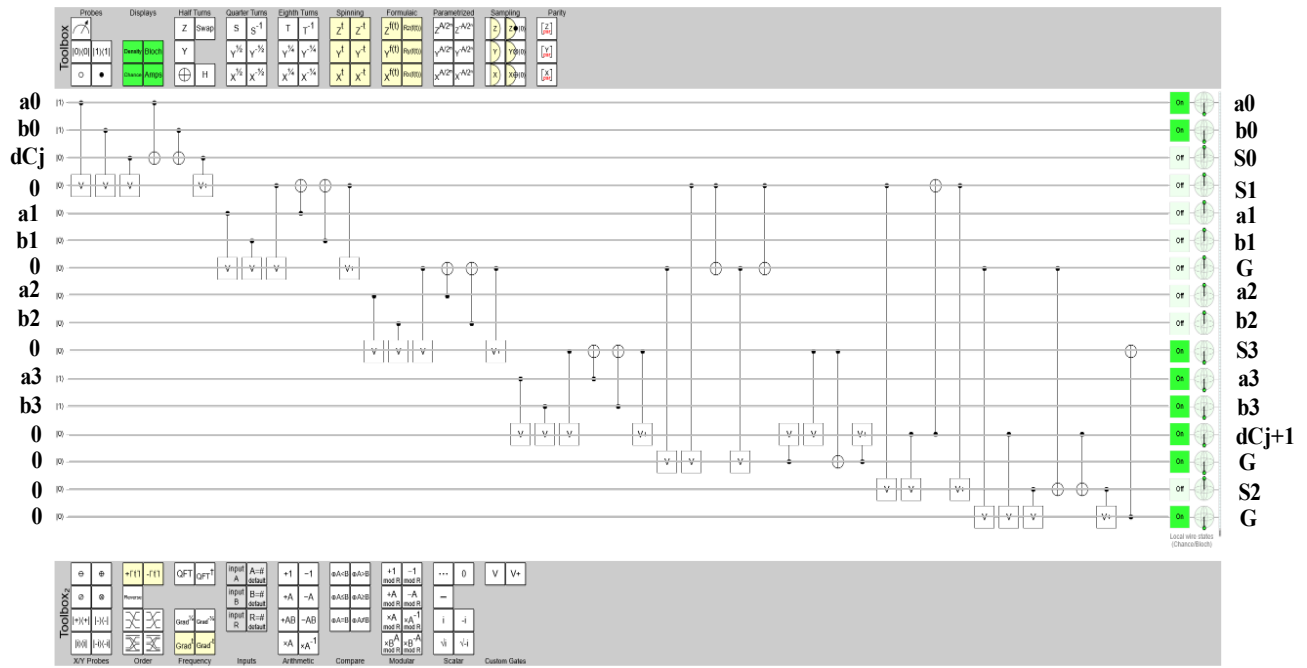

**Figure 8**. The proposed PDFA block for 1-digit for $A = a_3 a_2 a_1 a_0 = "1001"$, $B = b_3 b_2 b_1 b_0 = "1001"$, and $dcj = '0'$, simulated in the Quirk online tool.

The *Dec-CSK* (Figure 6) is described in VHDL and verified through simulations conducted in ModelSim. To ensure the validation of the design, various input patterns are applied using a test-bench program. The simulation results confirm the correctness and functional validity of the proposed architecture. For instance, **Figure 9** presents the expected output waveforms generated in ModelSim for two representative input patterns: "A=B=88888889", and "A=88888889", "B=11111111", which are some of the worst-case scenarios in 8-digit BCD addition. The simulation is done using a typical TSMC 90 nm CMOS technology process with Synopsys Design Compiler.

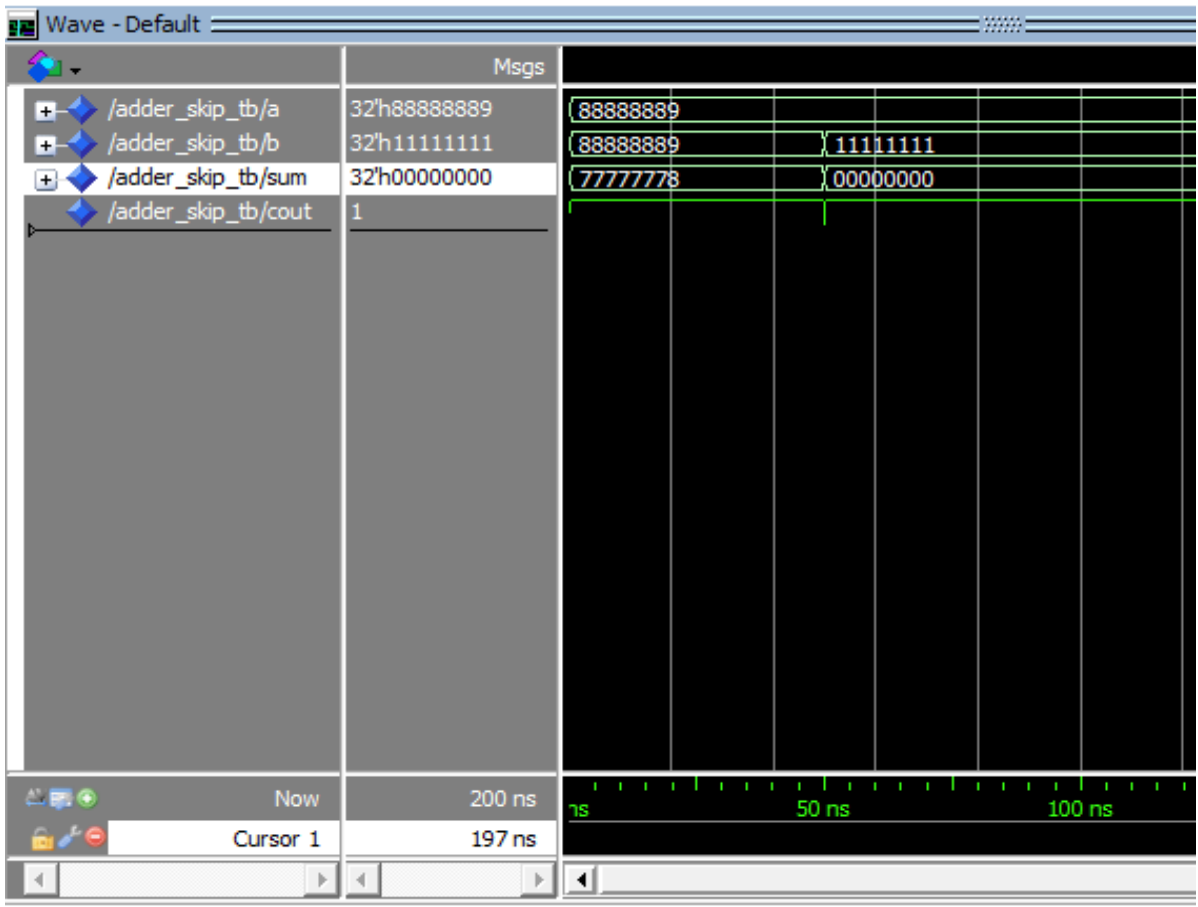

**Figure 9**. The output signals of the simulated *Dec-CSK* design for "A = B = 88888889", and "A = 88888889", "B = 11111111", using the Modelsim tool.

This part presents the evaluation and analysis of the proposed designs in comparison to existing counterparts. A comprehensive comparison of key features and performance metrics is summarized in Tables 2–5. Table 2 details the components of each proposed 1-digit BCD adder along with their corresponding figure of merit. Table 3 outlines the structural formulation of both the proposed and existing designs, highlighting their characteristics and comparing them based on key metrics such as CI, GO, QC, and circuit depth or delay. Tables 4 and 5 present an evaluation of the design parameters for a range of input sizes (N = 8, 16, 32, 64, 128, and 256 digits), providing a comprehensive foundation for analyzing scalability. Improvement percentages are calculated by comparing the proposed designs against each of the existing designs across various input sizes. The average of these improvements is then reported as the overall enhancement of the proposed designs relative to each counterpart in this section.

The two proposed BCD adders, each consisting of three primary components (as detailed in Section 3), with their quantum parameters, Gate Count (GC), CI, GO, QC, and Delay, are summarized in Table 2.

**Table 2.** Analysis of the proposed reversible BCD adder designs.

| Design | parameters | Initial Addition | Invalid Result Detection | Correction Mechanism | Total |
|---|---|---|---|---|---|
| **The Proposed *Dec-RCA* 1-digit BCD adder** | GC | 4 (4 HNG) | 3 (FG, PG, BJN) | 3 (PG, HNG, NOT) | 10 |
| | CI | 4 | 2 | 2 | 8 |
| | GO | 0 | 1 | 3 | 4 |
| | QC | 24 | 10 | 11 | 45 |
| | Delay | 20 Δ | 5 Δ | 10 Δ | 35 Δ |
| **The Proposed *Dec-CSK* 1-digit BCD adder** | GC | 4 (4 HNG) | 11 (4PG, 4FG, 2M-F, 1DFG) | 3 (PG, HNG, NOT) | 18 |
| | CI | 4 | 4 | 2 | 10 |
| | GO | 0 | 9 | 3 | 12 |
| | QC | 24 | 30 | 11 | 65 |
| | Delay | 20 Δ | 5 Δ | 10 Δ | 35 Δ |

**Table 3.** Comparison cost of the proposed and previous works of the reversible N-digit BCD adder [10-15]

| Design | CI | GO | QC | Delay |
|---|---|---|---|---|
| [10]-design1 | 11N | 16N | 58N | 40N |
| [10]-design2 | 12N | 17N | 75N | 40N |
| [11]-design1 | 2N | 2N -1 | 88N | 73N |
| [11]-design2 | N | N-1 | 70N | 57N |
| [12] | 17N | 22N | 81N | 54N* |
| [13] | 19N | 24N | 88N | 62N* |
| [14] | 7N | 7N | 56N | 40N |
| [15] | 10N | 14N | 52N | 31N |
| *Dec-RCA* | 8N | 4N | 45N | 25N+10 |
| *Dec-CSK* | 10N | 12N | 65N | 5N+40 |

*These values are reported by [10].

Table 3 presents the formulas for various quantum metrics, including quantum cost (QC), delay, and other performance parameters, which we calculated and generalized for different sizes. $N$. By generalizing these formulas, we provide a systematic framework to evaluate the proposed quantum designs across varying scales. The results in the table clearly show that our proposed designs achieve significant improvements in key metrics. In particular, they demonstrate lower quantum cost and reduced delay compared to existing designs, highlighting the efficiency and scalability of the proposed approach for larger quantum circuits. This analysis confirms that the proposed architectures offer both optimized resource usage and faster computation, making them suitable for practical implementation in quantum arithmetic operations.

As mentioned before, this work uses a PDFA block that applies as the fundamental unit for a two-BCD addition structure. In the *Dec-RCA* approach, this block is utilized in a ripple and straight-chain configuration. The *Dec-RCA* design features a simple structure with an efficient quantum cost. In contrast, the *Dec-CSK* design modifies the *Dec-RCA* architecture and incorporates a speed-up technique to reduce the critical path delay. Although this approach introduces some hardware overhead, it significantly decreases the overall delay. According to Tables 4 and 5, the *Dec-RCA* design achieves an average improvement of 30.77% in quantum cost compared to *Dec-CSK*, while the faster proposed design (*Dec-CSK*) achieves an average improvement of approximately 74% in delay compared to *Dec-RCA*.

According to the results in Tables 3- 5, Thabah et al. [10] introduced two BCD adder designs. The first design in [10] employed BJN and TG gates in the SCL section, while the second used other reversible gates, such as the PG. Our proposed *Dec-RCA* design adopts a similar approach by utilizing the BJN and PG blocks; however, it further enhances performance by integrating an efficient PG structure in the second adder stage. This modification results in superior performance across all key figure of merit compared to [10]. The proposed *Dec-CSK* design demonstrates significant improvements in QC and delay when compared to the best-performing design in [10]. The proposed *Dec-CSK* design achieves an average improvement of 83.73% in speed, while the proposed *Dec-RCA* design attains 22.41% enhancement in QC compared to [10] across input sizes ranging from 8 to 256 digits.

Thapliyal [11] introduced a new 4-bit binary quantum adder and applied it in two BCD adder approaches: a designed BCD adder and a BCD converter. Although the primary focus was on minimizing the number of CI, and GO, it led to increased QC and delay. In contrast, our two proposed designs aim to achieve a balanced trade-off between delay and quantum metrics. In the proposed *Dec-RCA* design, the RCA structure is employed to optimize the quantum cost of the circuit, based on Tables 4 and 5, achieving an average improvement of 35.71% for input sizes ranging from 8 to 256 digits compared to the reference design [11]. In the proposed *Dec-CSK* design, a speedup technique based on the CSK method is implemented to minimize the critical path delay, achieving an average improvement of 88.35% across input sizes ranging from 8 to 256 digits, while simultaneously enhancing the overall quantum cost efficiency. These results confirm that the proposed architectures offer a balanced and scalable trade-off between speed and resource utilization.

In [12], a reversible gate was proposed and employed in a BCD adder design; however, the overall implementation exhibited limited efficiency. In contrast, the proposed architectures in this work propagate the carry output to the next digit before the correction step. This modification effectively reduces the carry propagation delay and enables parallel execution of the MSB computation and sum correction at each stage. As a result, the proposed *Dec-CSK* design achieves an average improvement of 87.70% in speed, while the proposed *Dec-RCA* design attains an average enhancement of 44.44% in QC compared to [12] across input sizes ranging from 8 to 256 digits based on Tables 4 and 5.

In [13], a DFA block was introduced using reversible gates and applied to the construction of a BCD adder. However, the overall design demonstrated limited efficiency in quantum metrics. Based on Tables 4 and 5, the proposed *Dec-RCA* design achieves an average improvement of 48.86% in quantum cost, and the proposed *Dec-CSK* design reaches an average enhancement of 89.48% in delay across input sizes ranging from 8 to 256 digits compared to [13]. As a result, both of our proposed designs outperform the approach in [13] across all quantum metrics and delay.

In [14], Nagamani introduced a DFA block aimed at reducing the number of qubits and applied it to the design of a BCD adder. Their approach employed uncomputation and input regeneration techniques to minimize garbage outputs. In contrast, our proposed designs focus on optimizing QC and delay while maintaining a balanced trade-off among key quantum metrics. Based on Tables 4 and 5, the proposed *Dec-RCA* design yields a 19.64% improvement in QC, whereas the proposed *Dec-CSK* design achieves an 83.90% enhancement in delay on average, compared with [14] across input sizes ranging from 8 to 256 digits.

Haghparast [15] introduced a parity-preserving reversible gate, which employed to construct a new DFA block in the RCA chain. Due to applying the parity feature, the design increased the QC and delay. Based on Tables 4 and 5, the *Dec-RCA* design, employing the RCA structure, achieves a 13.46% improvement in QC, whereas the *Dec-CSK* design provides an average enhancement of 78.74% in delay compared to [15] for input sizes ranging from 8 to 256 digits.

**Table 4**. Quantum Cost comparison of N-digit proposed designs and existing reversible BCD adders [10–15].

| digit | [10] | [11] | [12] | [13] | [14] | [15] | Dec-RCA | Dec-CSK | % Impr Dec-RCA | % Impr Dec-CSK |
|---|---|---|---|---|---|---|---|---|---|---|
| 8 | 464 | 560 | 648 | 704 | 448 | 416 | 360 | 520 | 30.75 | −0.02 |
| 16 | 928 | 1120 | 1296 | 1408 | 896 | 832 | 720 | 1040 | 30.75 | −0.02 |
| 32 | 1856 | 2240 | 2592 | 2816 | 1792 | 1664 | 1440 | 2080 | 30.75 | −0.02 |
| 64 | 3712 | 4480 | 5184 | 5632 | 3584 | 3328 | 2880 | 4160 | 30.75 | −0.02 |
| 128 | 7424 | 8960 | 10368 | 11264 | 7168 | 6656 | 5760 | 8320 | 30.75 | −0.02 |
| 256 | 14848 | 17920 | 20736 | 22528 | 14336 | 13312 | 11520 | 16640 | 30.75 | −0.02 |
| **Total average percentage improvement (Reduction)** | | | | | | | | | 30.75 | −0.02 |

In Tables 4 and 5, the overall improvement percentage for the specified criteria compared to the counterparts [10–15] is obtained by averaging the improvement values listed in the "% Impr Dec-RCA" and "% Impr Dec-CSK" columns. For example, the total improvement (reduction) of the "*Dec-RCA*" in terms of QC is equal to 30.75% which is calculated by considering the average of the "% Impr *Dec-RCA*" column in Table 4 for all sizes from 8- to 256-digits. Similarly, the total percentage improvement (reduction) in delay for the "Dec-RCA" compared to the other counterparts [10-15] is calculated based on the aforementioned explanations for all sizes (8-256 digits) and is found to be 43.07%. As another example, the total improvement of the "*Dec-CSK*" in terms of QC is equal to 2% enhancement of this criterion compared to other counterparts (which is calculated by considering the average of the "% Imp *Dec-CSK*" column in Table 4 for all sizes from 8- to 256-digits). Similarly, the total percentage improvement (reduction) in delay for the "Dec-CSK" compared to the other counterparts [10-15] is calculated based on the aforementioned explanations for all sizes (8-256 digits) and is found to be 85.12%.

**Table 5**. Quantum delay comparison of N-digit proposed designs and existing reversible BCD adders [10–15].

| digit | [10] | [11] | [12] | [13] | [14] | [15] | Dec-RCA | Dec-CSK | % Impr Dec-RCA | % Impr Dec-CSK |
|---|---|---|---|---|---|---|---|---|---|---|
| 8 | 320 | 456 | 432 | 496 | 320 | 248 | 210 | 80 | 41.18 | 77.59 |
| 16 | 640 | 912 | 864 | 992 | 640 | 496 | 410 | 120 | 42.55 | 83.19 |
| 32 | 1280 | 1824 | 1728 | 1984 | 1280 | 992 | 810 | 200 | 43.28 | 85.99 |
| 64 | 2560 | 3648 | 3456 | 3968 | 2560 | 1984 | 1610 | 360 | 43.63 | 87.40 |
| 128 | 5120 | 7296 | 6912 | 7936 | 5120 | 3968 | 3210 | 680 | 43.89 | 88.10 |
| 256 | 10240 | 14592 | 13824 | 15872 | 10240 | 7936 | 6410 | 1320 | 43.92 | 88.45 |
| **Total average percentage improvement (Reduction)** | | | | | | | | | 43.07 | 85.12 |

As mentioned before, this paper seeks to balance conflicting objectives, such as minimizing delay and reducing quantum resources, in the design of decimal adders, with particular emphasis on improving computational speed. Consequently, two primary performance metrics are evaluated: delay and QC. To offer a thorough assessment of this trade-off, **Figure 10** depicts the trade-off between QC and delay for three input sizes (16, 32, and 64 digits), comparing the two proposed architectures with existing designs. The graphs apply the concept of Pareto optimality [35] to highlight the Pareto frontier, illustrated as the Pareto optimal curve within the scatter plots.

This graph presents the results of the BCD adder for three different sizes: blue dots represent the delay versus quantum cost of 16-digit adders, orange dots indicate the delay versus quantum cost of 32-digit adders, and green dots correspond to the delay versus quantum cost of 64-digit adders. The blue, orange, and green curves represent the optimal curves for 16-, 32-, and 64-digit, respectively. Both the *Dec-RCA* and *Dec-CSK* proposed designs lie on the Pareto optimal curve across all input sizes. For the 32-digit case, *Dec-RCA* and *Dec-CSK* achieve an optimal balance between delay and QC. The 32-digit proposed designs even outperform the designs in [11] and [13] developed for 16-digit adders. This performance advantage persists at larger scales as well. At 64 digits, the proposed designs continue to demonstrate superior efficiency compared to [11] and [13] even at smaller input sizes. According to Figure 10, *Dec-RCA* is a cost-efficient structure that achieves a low quantum cost. It is positioned in the upper region of the Pareto frontier, near the Y-axis. The *Dec-CSK* is optimized for delay and represents a high-speed structure. It is located in the lower region of the Pareto frontier, near the X-axis, across all three input sizes.

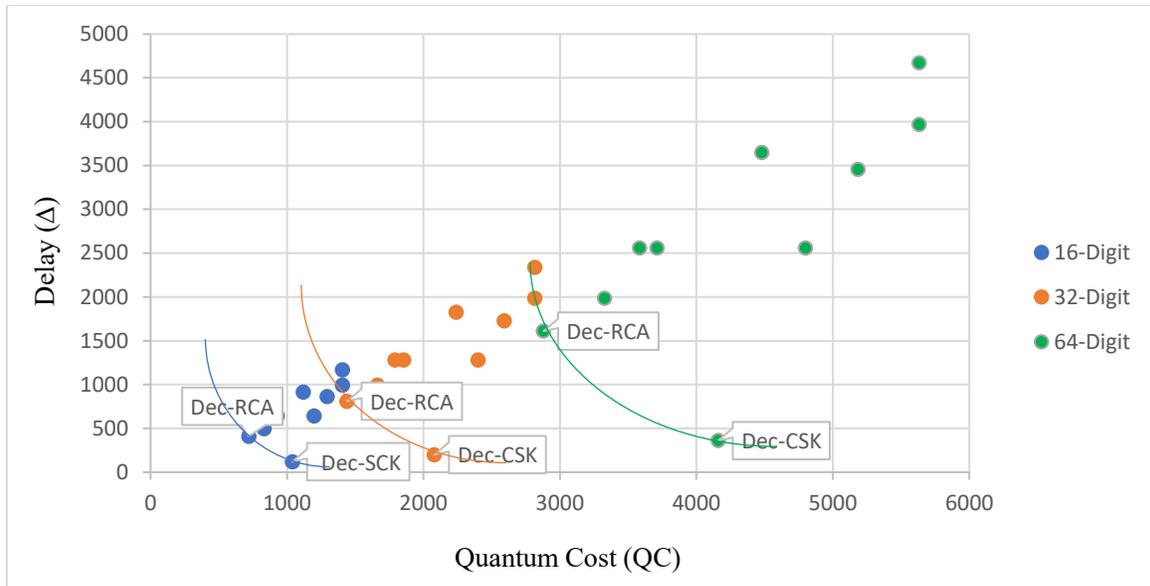

**Figure 10**. Trade-off graphs between a delay and Quantum Cost (QC) of the proposed and existing designs and their Pareto optimal curve.

BCD adders are essential components in various computational domains that require precise decimal arithmetic, such as digital signal processing, data analytics, and neural network architectures designed for decimal-based computation. In neural systems, for instance, a specialized layer utilizing BCD adders can be developed to process decimal inputs directly, ensuring accuracy in addition while maintaining compatibility with machine learning models. Beyond general-purpose arithmetic, the most critical domain for BCD adders is the financial and banking sector, where exact decimal handling is indispensable. These adders are used in a wide range of monetary operations, including deposits, withdrawals, fund transfers, interest and tax calculations, transaction fee computation, currency conversion, and accounting processes such as invoicing.

A representative example of practical BCD integration can be seen in the IBM z10 mainframe processor [36], which features a dedicated decimal unit employing BCD adders for high-precision decimal computation. This architecture supports direct execution of complex mathematical and statistical operations in decimal format, eliminating binary conversion and ensuring exact numerical results. Such precision is crucial in financial applications where even minor rounding errors can lead to inconsistencies. By natively supporting decimal arithmetic, the IBM z10 improves computational reliability and simplifies data conversion between machine and human-readable formats.

Building on this foundation, the proposed quantum BCD adders introduce a more scalable and efficient approach for future financial computation systems. Quantum BCD addition leverages quantum parallelism to execute large-scale decimal arithmetic operations at a higher speed and lower computational cost, enhancing both efficiency and precision. This capability makes it particularly valuable for real-time transaction verification, fraud detection, and large-scale financial analytics that demand exact decimal processing.

To demonstrate the practical potential of the proposed quantum BCD adder, a real-world dataset from the Kaggle platform [37] is analyzed. The dataset includes financial transactions, customer details, and card information from a banking institution across the 2010s decade. Using the BCD adder, total withdrawals per user across multiple cards are computed accurately without rounding or precision errors. Due to current limitations in quantum hardware and the restricted number of available qubits, the analysis is performed on 2,000 transactions, involving approximately 1,180 BCD addition operations. These experiments emphasize the importance of efficient quantum BCD circuit design, showing that optimization in quantum cost and delay can substantially improve computational performance and scalability in quantum-based financial systems.

## 6. Conclusion

The quantum and reversible paradigm integrates the fundamental principles of quantum mechanics with reversible computation to facilitate information-conserving and logically consistent processing. This conceptual framework provides a basis for the development of advanced computing architectures that enhance computational precision, scalability, and transparency. This paper introduces two novel BCD adders by applying the logical function of the decimal carry digit in a reversible structure. The proposed designs utilize a new correction and skip block combined with a speed-up technique to achieve an efficient trade-off between QC and delay. The *Dec-RCA* adder achieves a 30.75% average reduction in quantum cost compared to existing counterparts. The *Dec-CSK* design employs the CSK technique in decimal carry propagation, resulting in an 85.12% average improvement in delay. These designs have strong potential applications in the development of advanced financial quantum computing systems.

**Acknowledgement:**

This work is based upon research funded by the Iran National Science Foundation (INSF) under project No**.** 4041040.

**Author Contribution:**

Negin Mashayekhi contributed to conceptualization, methodology, validation, evaluation, formal analysis, investigation, data curation, writing, review and editing, and supervision. Mohammad Reza Reshadinezhad, Antonio Rubio, and Shekoofeh Moghimi contributed to formal analysis, investigation, resource, data curation, writing, review and editing, and supervision.